\documentclass{PoS}
\usepackage{url}
\title{Probing nuclear gluons with heavy quarks at EIC}
\ShortTitle{Probing nuclear gluons with heavy quarks at EIC}
\author{E.~Chudakov,$^a$ D.~Higinbotham,$^a$ Ch.~Hyde,$^b$ S.~Furletov,$^a$
Yu.~Furletova,$^a$ D.~Nguyen,$^{a,c}$ M.~Stratmann,$^d$ M.~Strikman,$^e$
\speaker{C.~Weiss},$^a$ R.~Yoshida$^a$ \\
\llap{$^a$} Jefferson Lab, Newport News, VA 23606, USA \\
\llap{$^b$} Old Dominion University, Norfolk, VA 23529, USA \\
\llap{$^c$} University of Virginia, Charlottesville, VA 22904, USA \\
\llap{$^d$} T\"ubingen University, 72076 T\"ubingen, Germany \\
\llap{$^e$} Pennsylvania State University, University Park, PA 16802, USA}
\abstract{We explore the feasibility of direct measurements of nuclear gluon densities
using heavy-quark production (open charm, beauty) at a future Electron-Ion Collider (EIC).
We focus on the regions $x > 0.3$ (EMC effect) and $x \sim 0.05$--$0.1$ 
(antishadowing), where the nuclear modifications of the gluon density offer insight into 
non-nucleonic degrees of freedom and the QCD structure of nucleon-nucleon interactions. 
We describe the charm production rates and momentum distributions in nuclear deep-inelastic 
scattering (DIS) at large $x_B$, and comment on the possible methods for charm 
reconstruction using next-generation detectors at the EIC 
($\pi / K$ identification, tracking, vertex detection).}
\FullConference{XXIV International Workshop on Deep-Inelastic Scattering and Related Subjects\\
11-15 April, 2016\\
DESY Hamburg, Germany}
\begin{document}
Nuclear parton densities (PDFs) describe the basic quark-gluon structure of the nucleus 
in QCD. They represent the expectation value of the leading-twist QCD operators in the 
nuclear ground state and can be measured in high-energy, high-momentum transfer processes 
such as DIS or dilepton production. The comparison of the nuclear PDFs 
to that of a system of unbound protons and neutrons ($A = Z + N$) offers unique insight 
into the QCD structure of nucleon interactions and the microscopic origin of 
nuclear binding. Distinct dynamical mechanisms are expected to cause nuclear modifications in 
different regions of $x$: modified single-nucleon structure and non-nucleonic degrees 
of freedom in nuclei ($x > 0.3$), exchange interactions between nucleons ($x \sim 0.1$,
antishadowing), and the appearance of coherent gluon fields associated with multiple
nucleons ($x < 0.01$, shadowing) \cite{Frankfurt:1988nt}.\footnote{Here $x$ refers to the
light-cone momentum fraction of the parton relative to a nucleon in the nucleus ($0 < x < 1$).}
Suppression of the nuclear quark densities in the valence region $x > 0.3$ has been observed 
in inclusive DIS (EMC effect) and is the object of intense theoretical 
study \cite{Malace:2014uea}.
Some information on nuclear sea quarks is available from dilepton production.

\begin{figure}[b]
\parbox[c]{0.5\textwidth}{
\includegraphics[width=0.5\textwidth]{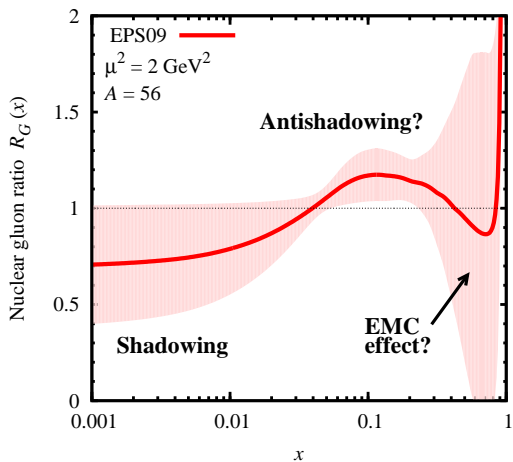}}
\hspace{0.05\textwidth} 
\parbox[c]{0.38\textwidth}{
\caption[]{The nuclear gluon density ratio 
$R_G(x, \mu^2) = G_A(x, \mu^2) / [A G_N(x, \mu^2)]$ and its uncertainty
at a scale $\mu^2 = 2$ GeV$^2$, obtained from the 
EPS09 analysis of nuclear PDFs \cite{Eskola:2009uj}. 
\label{fig:eps09}}}
\end{figure}

Much less is known about the nuclear modifications of gluons (see Fig.~\ref{fig:eps09}). 
Basic questions remain to be answered: (a) Is the nuclear gluon density suppressed 
at $x > 0.3$ (gluonic EMC effect)? This would provide insight into the change of 
quark-gluon configurations of the nucleon due to nuclear binding. (b) Are nuclear gluons 
enhanced at $x \sim 0.1$ (gluon antishadowing)? This would reveal the gluonic structure of 
nucleon-nucleon interactions at average distances in the nucleus. A recent theoretical
analysis \cite{Guzey:2013qza} of data in $J/\psi$ production in ultraperipheral 
$AA$ collisions at LHC \cite{Abbas:2013oua} confirms substantial gluon shadowing 
at $x < 0.01$, which suggests large compensating antishadowing at $x \sim 0.1$ to 
conserve the overall light-cone momentum carried by gluons.

The nuclear modifications of gluons at $x \gtrsim 0.1$ have so far been studied only 
indirectly, through the $Q^2$ dependence of inclusive nuclear DIS cross sections
(DGLAP evolution). Results could be improved by extending such measurements over a larger 
range of $Q^2$ and $W$, and separating transverse and longitudinal nuclear structure functions. 
Much more incisive could be measurements with probes coupling directly to gluons
at a fixed scale.
%
%
%

%
%
\begin{figure}
\begin{tabular}{ll}
\parbox[c]{0.33\textwidth}{\includegraphics[width=0.33\textwidth]{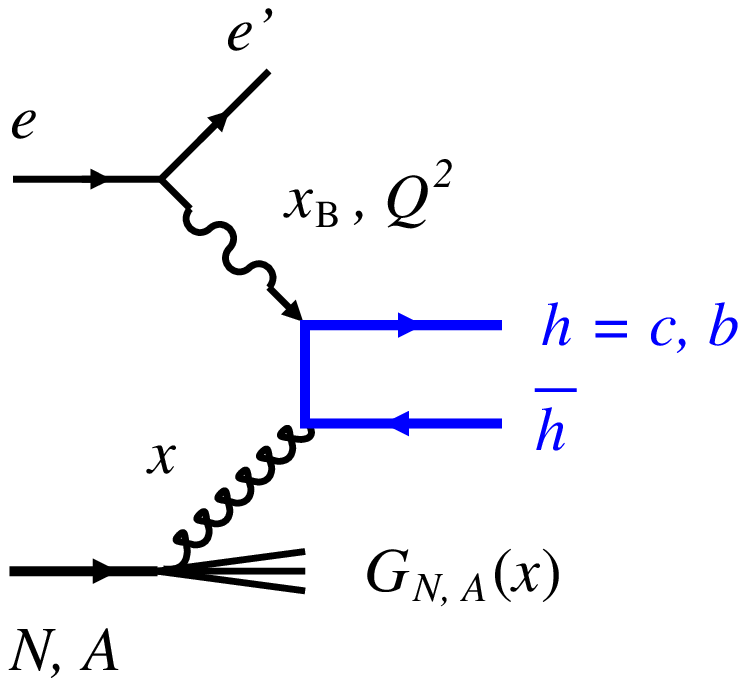}}
\hspace{0.05\textwidth}
&
\parbox[c]{0.48\textwidth}{\includegraphics[width=0.48\textwidth]{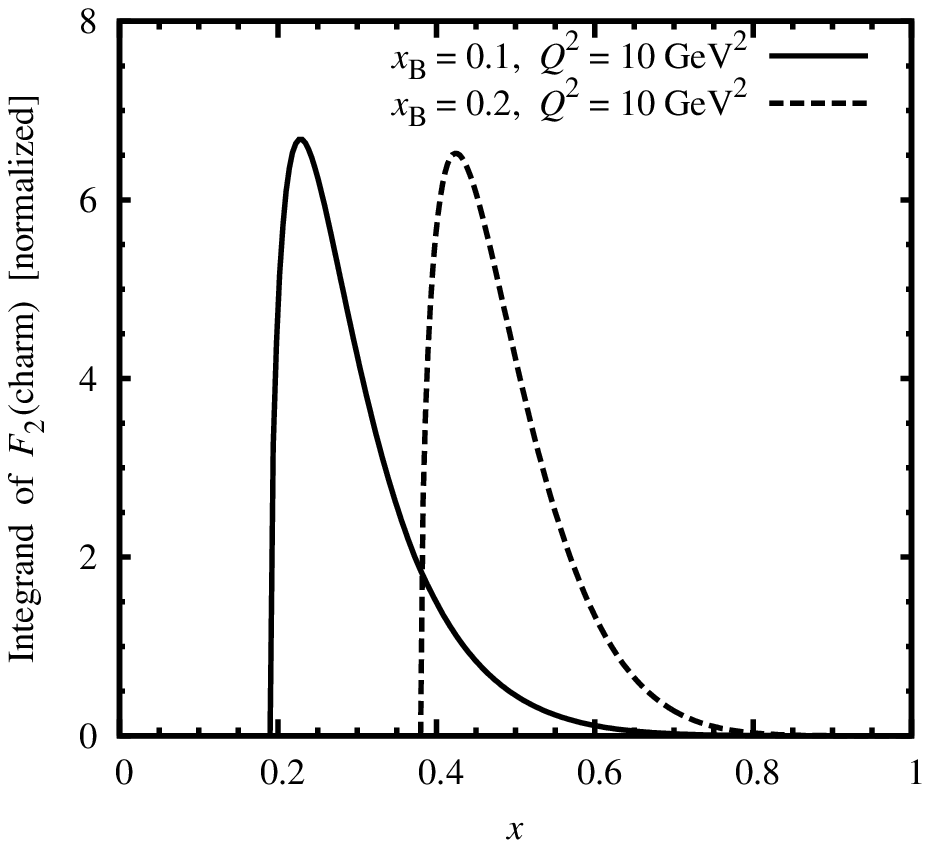}}
\\[-2ex]
{\small (a)} & {\small (b)}
\end{tabular}
\caption[]{(a) Heavy-quark production in DIS at LO (photon-gluon fusion).
(b) Distribution of gluon momentum fractions $x$ in the convolution integral 
defining the charm structure function $F_2\textrm{(charm)}$ at given $x_{\rm B}$ 
and $Q^2$ (values see plot). The distributions are shown normalized to unit integral.
\label{fig:hq}}
\end{figure}
Heavy quark production in DIS provides a direct probe of the gluon density in the target. 
At leading order (LO) in the perturbative QCD expansion the heavy quark
pair is produced through the photon--gluon fusion process (see Fig.~\ref{fig:hq}a). 
The heavy quark structure function is given by a convolution integral over the
gluon density extending over $x > ax_{\rm B}$, where $x_{\rm B}$ is the Bjorken variable
defined by the electron kinematics, and $a = 1 + 4 m_h^2/Q^2$ ($m_h$ is the
heavy quark mass). The process effectively probes the gluon density in a region 
of $x$ strongly localized above $x_{\rm B}$ (see Fig.~\ref{fig:hq}b), and at a scale 
$\mu^2 \approx 4 m_h^2$ \cite{Gluck:1993dpa}. Next-to-leading order (NLO) QCD 
corrections have been calculated \cite{Laenen:1992zk}, and the theoretical 
uncertainties have been quantified \cite{Baines:2006uw}. Extensive measurements of 
charm and beauty production have been performed at the HERA $ep$ collider at 
$x_{\rm B} < 0.01$, using various methods of charm/beauty identification 
(see below), and found good agreement with QCD predictions \cite{Aaron:2010ib}. 
Open charm production was also studied at the COMPASS $\mu N$ fixed-target experiment
\cite{Adolph:2012uz}.

The proposed Electron-Ion Collider (EIC) \cite{2015NSAC,EIC-designs} would enable the 
first direct measurements of nuclear gluons at intermediate and large $x$ using heavy 
quark probes and could qualitatively advance our understanding of the gluonic structure 
of nuclei. The electron-nucleon squared center-of-mass energy in the range 
$s_{eN} \equiv s_{eA}/A \sim$ 200--2000 GeV$^2$ would provide wide coverage 
in $Q^2$ in DIS at $x_{\rm B} > 0.01$. [We quote the $s_{eN}$ values for $eA$ collisions, 
which are lower than those for $ep$ by the factor $Z/A \approx 0.5 (0.4)$ for light (heavy)
nuclei.] The luminosity of the order $10^{34}$ cm$^{-2}$ s$^{-1}$ (per nucleon) would give 
reasonable charm production rates even at $x_{\rm B} \gtrsim 0.1$. The next-generation 
detectors ($\pi / K$ identification, tracking, vertex detection) would open up new methods 
of charm reconstruction with greater efficiency than those used at HERA. Here we report 
about results of an R\&D project \cite{LD1601} aiming to explore the feasibility of 
such measurements on nuclei and to quantify their physics impact (for a general overview 
of nuclear physics with EIC, see Refs.~\cite{Accardi:2011mz,Boer:2011fh}). 

%
%
\begin{figure}
\parbox[c]{0.64\textwidth}{
\includegraphics[width=0.64\textwidth]{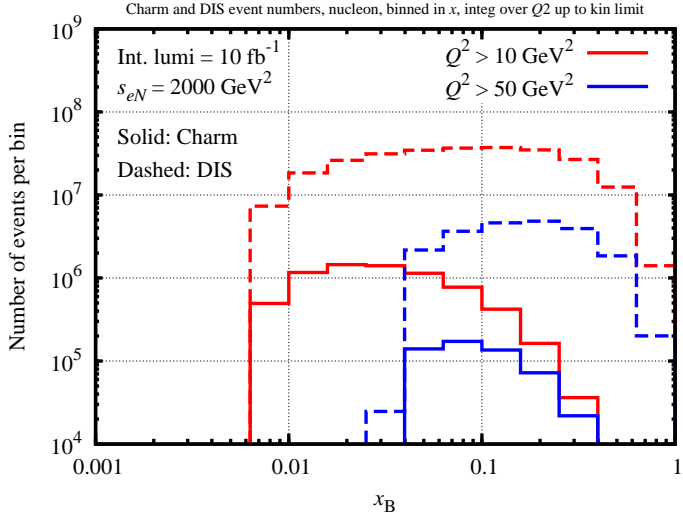}}
\hspace{0.02\textwidth} 
\parbox[c]{0.3\textwidth}{
\caption[]{Solid lines: Estimated number of charm events in DIS 
at EIC (CM energy $s_{eN} = 2000$ GeV$^2$, 
integrated nucleon luminosity 10 fb$^{-1}$). The bins in $x_{\rm B}$ are
5 per decade as indicated on the plot. $Q^2$ is integrated from the
lower limit indicated (10 or 50 GeV$^2$) to the kinematic limit.
Dashed lines: Total number of DIS events in the same bins
\label{fig:rates}}}
\end{figure}
Charm production rates in nuclear DIS at EIC have been estimated using QCD
expressions and the HVQDIS code \cite{Harris:1997zq} (see Fig.~\ref{fig:rates}). 
One observes that:
(a) the charm rates drop rapidly above $x_{\rm B} \sim 0.1$, due to the drop 
of the gluon density; (b) charm rates of few $\times 10^{5}$
can be achieved at $x_{\rm B} \sim 0.1$ with 10 fb$^{-1}$ integrated luminosity; 
(c) the fraction of DIS events with charm production
at $x_{\rm B} \sim 0.1$ 
is $\sim 1\%$ at $Q^2 > 10$ GeV$^2$ ($\sim 3\%$ at $Q^2 > 50$ GeV$^2$). Thus   
large--$x$ charm samples of $O(10^4)$ could be obtained if charm events could be 
identified with an overall efficiency of $\sim 10\%$.

At an EIC with ion momenta $\sim$ few 10 GeV (per nucleon) the charm quarks produced 
in DIS at $x_{\rm B} \sim 0.1$ typically emerge with large angles in the lab frame, 
and carry momenta of $\sim$ several GeV. (The actual distribution is determined by 
the transformation from the collinear frame, where the momentum transfer $\vec{q}$ 
and the ion momentum $\vec{p}_A$ are collinear, to the lab frame and exhibits
a complex dependence on $x_{\rm B}$ and $Q^2$.) The charm angular and momentum 
distributions are imparted on the produced $D$ mesons and their final decay 
hadrons ($\pi, K$). Detection capabilities for these hadrons need to be provided in 
the relevant angle and momentum ranges. An advantage of the moderate ion beam 
energies $\sim$ few 10 GeV (per nucleon) is that the hadrons are produced at large 
angles and momenta $\sim$ few GeV, where good particle identification (PID) 
can be performed.

\begin{figure}[t]
\includegraphics[width=0.95\textwidth]{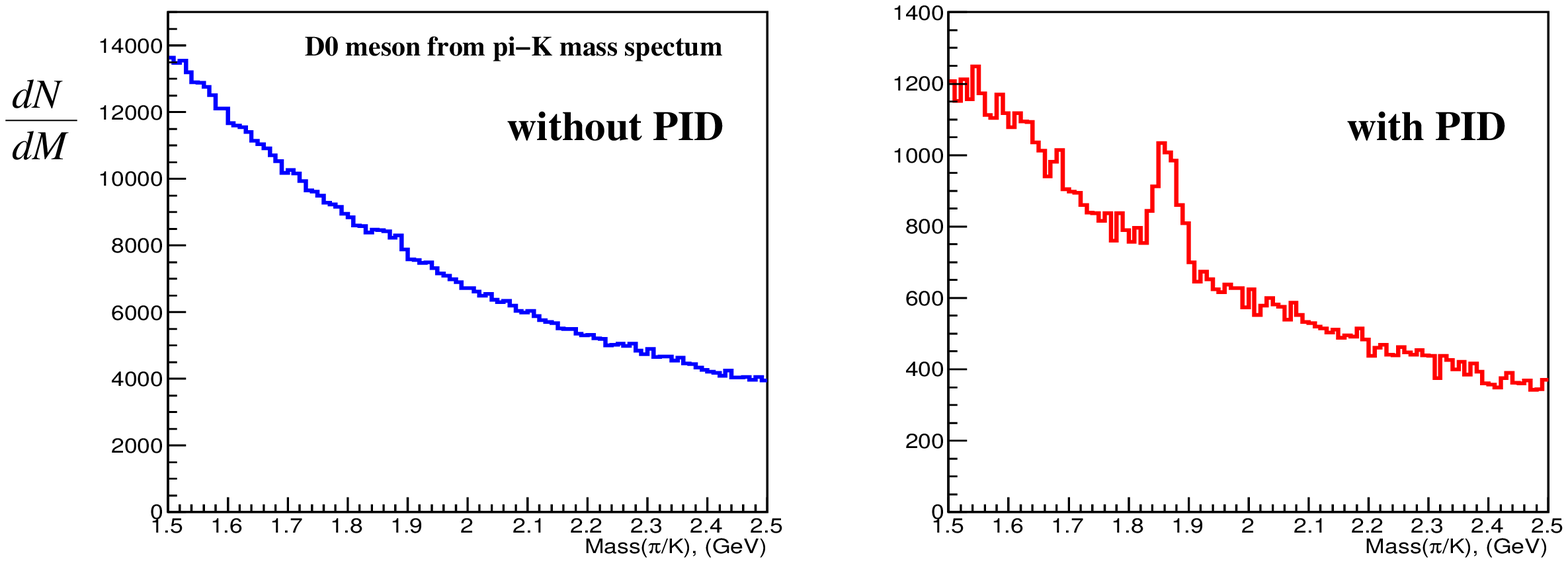}
\caption[]{Impact of PID on $D^0$ meson reconstruction from 
charged $\pi/K$ decays. The plots show the invariant mass spectrum 
of charged meson pairs, $dN/dM$, in a sample of charm events without and with PID 
(arbitrary normalization, no DIS background included). The peak represents
the $D^0 \rightarrow K^- \pi^+$ decay.
\label{fig:pid}}
\end{figure}

Charm events are identified by reconstructing the $D$ mesons that
are produced by charm quark fragmentation and subsequently decay into $\pi$ and $K$.
Experiments at HERA \cite{Aaron:2010ib} made extensive use of the $D^\ast$ channel, 
which exhibits a distinctive two-step decay  $D^{\ast +} \rightarrow D^0 \pi^+$(slow), 
$D^0 \rightarrow K^- \pi^+$, and can be reconstructed without PID or vertex detection.
However, this channel offers an overall reconstruction efficiency of only $\lesssim 1\%$
(given by the product of the fragmentation function and the branching ratios)
and will likely not be sufficient for nuclear gluon measurements at $x_{\rm B} \sim 0.1$.
The EIC detector will provide vastly improved PID capabilities, especially for
charged $\pi/K$ separation, and allows one to use other decay channels to reconstruct
$D$ mesons through charged pi/K tracks (see Fig.~\ref{fig:pid}). A survey of 
decay channels shows that this method could potentially permit charm reconstruction 
with an efficiency of $\sim 10\%$, which would significantly expand the physics 
reach at large $x_B$. Simulating charm reconstruction with this method and optimizing 
its performance are objects of on-going R\&D \cite{LD1601}.
%
%

Reconstruction of the displaced decay vertex of the $D$ meson can substantially 
improve the signal/background ratio in charm/beauty reconstruction, by eliminating much 
of the combinatorial background. However, the method reduces the overall charm/beauty
reconstruction efficiency, because it rejects events with a short decay length. 
Vertex detection was/is extensively used in high-energy experiments (HERA, LHC), 
where the charm production rates are large and the decay lengths are boosted by the 
large $D$ meson momenta. In DIS at EIC at $x_{\rm B} \sim 0.1$ the charm cross section will 
be $\sim 1\%$ of the total cross section, so that it is imperative to maximize the overall 
efficiency of charm reconstruction. At the same time the vertex displacements will
be smaller than in the high-energy experiments. The benefits of vertex detection in 
this context need to be explored. 

Another possible strategy for large--$x$ gluon measurements with charm is to focus 
on exceptional $c\bar c$ pairs with large transverse momenta $p_T \sim Q$.
While they are produced with a small cross section, such configurations represent 
a very distinctive final state that is practically free from hadronic background.
Whether nuclear ratio measurements would be feasible with such final states
is a topic for further R\&D \cite{LD1601}.

Measurements of nuclear ratios of charm production with EIC require good control
of the relative nuclear luminosity of the different ion beams. One possible method 
is to normalize the luminosity through measurement of the inclusive nuclear DIS structure 
function ratio $F_2^A/F_2^D$ at $x_B \sim 0.2-0.3$ and $Q^2 \sim \textrm{few GeV}^2$, 
where the nuclear modification was measured in fixed-target experiments and is known 
to be very small (``double ratio method'').

It is also necessary to analyze the theoretical uncertainties associated with 
nuclear charm production measurements. The uncertainties related to the QCD 
subprocess (higher-order corrections, effective scale) cancel when taking nuclear ratios,
making such measurements more robust than those of the absolute gluon density. 
The effects of nuclear final-state interactions on the observed
$D$--meson spectrum can be separated from initial-state modifications of the
nuclear gluon density by using the different $A$--dependence of the two mechanisms.
Finally, the impact of the charm production pseudodata on the nuclear PDFs can be 
quantified using reweighting methods \cite{Paukkunen:2014zia}.

In sum, direct measurements of nuclear gluons at $x > 0.1$ could significantly
advance our understanding of nucleon interactions in QCD. A medium-energy EIC with
ion beam energies $\sim$ several 10 GeV (per nucleon) and luminosity $10^{34}$ cm$^{-2}$
s$^{-1}$ is ideally suited for this purpose. Large-$x$ gluon measurements will be limited
by rates (luminosity) and the efficiency of charm reconstruction. An overall efficiency 
of up to $\sim 10\%$ could possibly be achieved using the PID capabilities of the 
EIC detector. Further R\&D is needed to demonstrate the method \cite{LD1601}.

This material is based upon work supported by the U.S.~Department of Energy, 
Office of Science, Office of Nuclear Physics under contract DE-AC05-06OR23177.

\end{document}